\documentclass[prd,preprint,a4paper,superscriptaddress,nofootinbib,11pt]{revtex4}
\usepackage{amsmath,amsfonts,amssymb}
\usepackage{txfonts}
\usepackage[T1]{fontenc}
\renewcommand{\d}{\mathrm{d}}
\begin{document}

%%%%%%%%%%%%%%%NASLOVNA%%%%%%%%%%%%%%%%%%%%%%%%%%%%%%%%%%%
%%%%%%%%%%%%%%%%%%%%%%%%%%%%%%%%%%%%%%%%%%%%%%%%%%%%%%%%%%%%
\begin{center}
{\bf  \Large Differential forms  and $\kappa$-Minkowski spacetime from extended twist\\}
 
 \bigskip
\bigskip

Tajron Juri\'c  {\footnote{e-mail:e-mail: tjuric@irb.hr}} \\  
Rudjer Bo\v{s}kovi\'c Institute, Bijeni\v cka  c.54, HR-10002 Zagreb,
Croatia \\[3mm]

Stjepan Meljanac {\footnote{e-mail: meljanac@irb.hr}},
 \\  
Rudjer Bo\v{s}kovi\'c Institute, Bijeni\v cka  c.54, HR-10002 Zagreb,
Croatia \\[3mm] 
 
Rina  \v{S}trajn {\footnote{e-mail: r.strajn@jacobs-university.de}},
\\
Jacobs University Bremen, 28759 Bremen, Germany\\[3mm]

\end{center}
\setcounter{page}{1}

%%%%%%%%%%%%%%%%%%%%%%%%%%%%%%%% ABSTRACT %%%%%%%%%%%%%%%%%%%%%%%%%%%%%%%%%%%%

{
 We analyze bicovariant differential calculus on $\kappa$-Minkowski spacetime. It is shown that corresponding Lorentz generators and noncommutative coordinates compatible with bicovariant calculus cannot be realized in terms of
commutative coordinates and momenta. Furthermore, $\kappa$-Minkowski space and NC forms are constructed by twist related to a bicrossproduct basis. It is pointed out that the consistency condition is not satisfied. We present the construction of $\kappa$-deformed coordinates and forms (super-Heisenberg algebra) using extended twist.  It is compatible with bicovariant differential calculus with $\kappa$-deformed $\mathfrak{igl}(4)$-Hopf algebra. The extended twist leading to $\kappa$-Poincar\'{e}-Hopf algebra is also discussed.}


%%%%%%%%%%%%%%%%%%%%%%%%%%%%%%%%%%%%%%%%%%%%%%%%%%%%%%%%%%%%%%%%%%%%%%%%%%%%%%%%%%
%%%%%%%%%%%%%%%%%%%%%%%%%%%%%%%%%%%%%%%%%%%%%%%%%%%%%%%%%%%%%%%%%%%%%%%%%%%

\section{Introduction}
The structure of spacetime at very high energies (Planck scale lengths) is still unknown and it is believed that, at these energies, gravity effects become significant and we need to abandon the notion of smooth and continuous spacetime. Among many attempts to find a suitable model for unifying quantum field theory and gravity, one of the ideas that emerged is that of noncommutative spaces \cite{Connes}-\cite{Aschieri}. Authors inclined to this idea have followed different approaches and considered different types of noncommutative (NC) spaces, where the concept of invoking  twisted Poincar\'{e} symmetry of the
algebra of functions on a Minkowski spacetime using twist operator is the most elaborated one \cite{chaichan}.
 In formulating field theories on NC spaces, differential calculus  plays an essential role. The requirement that this differential calculus is bicovariant and also covariant under the expected group of symmetries leads to some problems.

One widely researched type of NC space, which is also the object of consideration in this letter, is the $\kappa$-Minkowski space \cite{Lukierski-1}-\cite{Gumesa}. This space is a Lie algebra type of deformation of the Minkowski spacetime and here the deformation parameter $\kappa$ is usually interpreted as the Planck mass or the quantum gravity scale. $\kappa$-Minkowski space is also related to  doubly special relativity \cite{Amelino-Camelia-1}-\cite{Kowalski-Glikman-2}. For each NC space there is a corresponding symmetry algebra. In the case of $\kappa$-Minkowski space, the symmetry algebra is a deformation of the Poincar\'{e} algebra, known as the $\kappa$-Poincar\'{e} algebra. The $\kappa$-Poincar\'{e} algebra is also an example of a Hopf algebra. Some of the results of pursuing this line of research are, e.g., the construction of quantum field theories \cite{klm00}-\cite{mstw11}, electrodynamics \cite{h}-\cite{jonke}, considerations of quantum gravity effects \cite{bgmp10}-\cite{hajume} and the modification of particle statistics \cite{kappaSt}-\cite{Gumesa} on $\kappa$-Minkowski space.

Regarding the problem of differential calculus on $\kappa$-Minkowski space, Sitarz has shown \cite{Sitarz} that in order to obtain bicovariant differential calculus, which is also Lorentz covariant, one has to introduce an extra cotangent direction. While Sitarz considered 3+1 dimensional space (and developed five dimensional differential calculus), Gonera et al. generalized this work to $n$ dimensions in Ref. \cite{Gonera}. Another attempt to deal with this issue was made  in \cite{klry08} by the Abelian twist deformation of $\text{U}[\mathfrak{igl}(4,\mathbb{R})]$.  Bu et al. in \cite{Bu-Kim} extended the Poincar\'{e} algebra with the dilatation operator and constructed a four dimensional differential algebra on the $\kappa$-Minkowski space using a Jordanian twist of the Weyl algebra. Differential algebras of classical dimensions were also constructed in \cite{Meljanac-1} and \cite{Meljanac-2}, from the action of a deformed exterior derivative.

In \cite{MKJj} the authors have constructed two families of differential algebras of classical dimensions on the $\kappa$-Minkowski space, using realizations of the generators as formal power series in a Weyl superalgebra. In this approach, the realization of the Lorentz algebra is also modified, with the addition of Grassmann-type variables. As a consequence, generators of the Lorentz algebra act covariantly on one-forms, without the need to introduce an extra cotangent direction. The action is also covariant if restricted to the $\kappa$-Minkowski space. However, one loses Lorentz covariance when considering forms of order higher than one.

Our motivation in this letter is to unify $\kappa$-Minkowski spacetime, $\kappa$-Poincare algebra and differential forms. We embed them into $\kappa$-deformed super-Heisenberg algebra related to bicrossproduct basis. Using extended twist, we construct a smooth mapping between $\kappa$-deformed super-Heisenberg algebra and super-Heisenberg algebra. We present an extended realization for $\kappa$-deformed coordinates, Lorentz generators, and exterior derivative compatible with Lorentz covariance condition.

In section II, super-Heisenberg algebra is described. In section III, realization of $\kappa$-Minkowski space and $\kappa$-Poincare algebra related to bicrossproduct basis is given. In section IV, bicovariant differential calculus is analyzed.  It is pointed out that there does not exist a realization of Lorentz generators and NC coordinates compatible with bicovariant calculus in terms of commutative coordinates and momenta. In section V, $\kappa$-Minkowski space and NC forms are constructed by twist related to bicrossproduct basis. It is shown that the consistency condition is not satisfied. In section VI,  we present our main construction of $\kappa$-deformed super-Heisenberg algebra using extended twist. Extended realizations for Lorentz generators and exterior derivative invariant under $\mathfrak{igl}(4)$-Hopf algebra are presented. Finally, in section VII we outline the construction of Lorentz generators, exterior derivative and one-forms for bicovariant calculus compatible with $\kappa$-Poincar\'{e}-Hopf algebra.

\section{Super-Heisenberg algebra}
In the undeformed case we consider spacetime coordinates $x_{\mu}$, derivatives $\partial_{\mu}\equiv \frac{\partial}{\partial x^\mu}$, one forms $\d x_{\mu}\equiv\xi_{\mu}$, and Grassmann derivatives $q_{\mu}\equiv\frac{\partial}{\partial \xi^{\mu}}$ satisfying the following (anti)commutation relations:
\begin{equation}\begin{split}\label{SH}
&[x_{\mu},x_{\nu}]=[\partial_{\mu},\partial_{\nu}]=0,\quad [\partial_{\mu},x_{\nu}]=\eta_{\mu\nu}, \\
&\{\xi_{\mu},\xi_{\nu}\}=\{q_{\mu},q_{\nu}\}=0, \quad  \{\xi_{\mu},q_{\nu}\}=\eta_{\mu\nu},\\
&[x_{\mu},\xi_{\nu}]=[x_{\mu},q_{\nu}]=[\partial_{\mu},\xi_{\nu}]=[\partial_{\mu},q_{\nu}]=0,\\
\end{split}\end{equation}
where $\mu=\left\{0,1,2,3\right\}$ and $\eta_{\mu\nu}=\text{diag}(-1,1,1,1)$. The algebra in (\ref{SH}) generates the undeformed super-Heisenberg algebra ${\cal SH}(x,\partial,\xi,q)$ i.e. superphase space. The exterior derivative is defined as $\d=\xi_{\alpha}\partial^{\alpha}$, so $\xi_{\mu}=[\d,x_{\mu}]$. 

We define the action 
$\triangleright : {\cal SH}(x,\partial,\xi,q)\mapsto {\cal SA}(x,\xi) $, where ${\cal SA}(x,\xi)\subset{\cal SH}(x,\partial,\xi,q)$. The super-Heisenberg algebra ${\cal SH}(x,\partial,\xi,q)$ can be written as ${\cal SH}={\cal SA}\;{\cal ST}$, where ${\cal ST}(\partial, q)\subset{\cal SH}(x,\partial,\xi,q)$. For any element $f(x,\xi)\in{\cal SA}(x,\xi)$ we have
\begin{equation}\begin{split}\label{djelovanje}
x_{\mu} &\triangleright f(x,\xi)=x_{\mu}f(x,\xi),\quad \xi_{\mu}\triangleright f(x,\xi)=\xi_{\mu}f(x,\xi) , \\
&\partial_{\mu}\triangleright f(x,\xi)=\frac{\partial f}{\partial x^{\mu}}, \quad  q_{\mu}\triangleright f(x,\xi)=\frac{\partial f}{\partial \xi^{\mu}}.\\
\end{split}\end{equation}

The coalgebra structure of ${\cal ST}(\partial, q)$ is defined by undeformed coproducts:
\begin{equation}\begin{split}\label{coproduct}
 \Delta_{0}\partial_{\mu}=\partial_{\mu}\otimes 1+&1\otimes \partial_{\mu},\quad  \Delta_{0}q_{\mu}=q_{\mu}\otimes 1+(-)^{\text{deg}}\otimes q_{\mu},\\
& \text{deg}=\xi_{\alpha}q^{\alpha}(\text{mod}2).\\
\end{split}\end{equation}
The coalgebra structure with antipode and counit is (undeformed) super-Hopf algebra. Let us mention that super-Heisenberg algebra ${\cal SH}$ has also super-Hopf-algebroid structure, which will be elaborated separately. The Hopf-algebroid structure of Heisenberg algebra was discussed in \cite{mali} \footnote{for Hopf-algebroid structure also  see \cite{Lu}, \cite{Bem} and \cite{rmatrix}}.

Now we introduce Lorentz generators $M_{\mu\nu}$:
\begin{equation}\label{lorentz}
[M_{\mu\nu},M_{\lambda\rho}]=\eta_{\nu\lambda}M_{\mu\rho}-\eta_{\mu\lambda}M_{\nu\rho}-\eta_{\nu\rho}M_{\mu\lambda}+\eta_{\mu\rho}M_{\nu\lambda},
\end{equation}
with the following undeformed coproduct:
\begin{equation}\label{M-co}
\Delta_{0} M_{\mu\nu}=M_{\mu\nu}\otimes 1+1\otimes M_{\mu\nu}
\end{equation}
 and action $\triangleright$ :
\begin{equation}\begin{split}\label{M-action}
 M_{\mu\nu}\triangleright x_{\lambda}=\eta_{\nu\lambda}x_{\mu}-&\eta_{\mu\lambda}x_{\nu},\quad M_{\mu\nu}\triangleright \xi_{\lambda}=\eta_{\nu\lambda}\xi_{\mu}-\eta_{\mu\lambda}\xi_{\nu}. \\
 &M_{\mu\nu}\triangleright 1=0\\
\end{split}\end{equation}
Using (\ref{M-co}) and (\ref{M-action}) we can derive the commutation relations 
\begin{equation}\label{Mxxi}
[M_{\mu\nu},x_{\lambda}]=\eta_{\nu\lambda}x_{\mu}-\eta_{\mu\lambda}x_{\nu}, \quad [M_{\mu\nu},\xi_{\lambda}]=\eta_{\nu\lambda}\xi_{\mu}-\eta_{\mu\lambda}\xi_{\nu},
\end{equation}
 so that $x_{\mu}$ and $\xi_{\mu}$ transform as vectors (the same holds for $\partial_{\mu}$ and $q_{\mu}$). 
 
 The Lorentz covariance condition
 \begin{equation}\begin{split}\label{lorentzcond}
 M_{\mu\nu}&\triangleright f(x,\xi)g(x,\xi)=m_{0}\big(\Delta_{0}M_{\mu\nu}\triangleright f\otimes g\big)\\
 &M_{\mu\nu}\triangleright \text{d}f(x,\xi)=\d(M_{\mu\nu}\triangleright f(x,\xi))\\
 \end{split}\end{equation}
(where $m_{0}$ is the multiplication map) implies
\begin{equation}\label{Md}
[M_{\mu\nu},\d]=0,
\end{equation}
where $\d f(x,\xi)=\d\triangleright f(x,\xi)=[\d, f(x,\xi)]\triangleright 1$ and $\d\triangleright 1=0$.
Note that the action $M_{\mu\nu}\triangleright f(x, \xi )$ in Eq. (\ref{lorentzcond}) is compatible with (\ref{M-action}), (\ref{Mxxi}) and
\begin{equation}
M_{\mu\nu}\triangleright f(x,\xi)g(x,\xi)=M_{\mu\nu}f(x,\xi)g(x,\xi)\triangleright 1
\end{equation}

 The realization for $M_{\mu\nu}$ in  ${\cal SH}(x,\partial,\xi,q)$ is
 \begin{equation}\label{realM}
 M_{\mu\nu}=x_{\mu}\partial_{\nu}-x_{\nu}\partial_{\mu}+\xi_{\mu}q_{\nu}-\xi_{\nu}q_{\mu}
 \end{equation}
and now it is easy to  verify Eqs.(\ref{lorentz}) - (\ref{Mxxi}). Note that the Lorentz generators without the Grassmann-part  ($\xi_{\mu}q_{\nu}-\xi_{\nu}q_{\mu}$) cannot satisfy the condition (\ref{Md}). Usually in differential geometry vector field $\text{v}=\text{v}_{\mu}\partial^{\mu}$ acts on a one-form $\xi_{\beta}=\d x_{\beta}$ as a Lie derivative $\mathcal{L}_{\text{v}}\xi_{\beta}=\d \mathcal{L}_{\text{v}}x_{\beta}=\d \text{v}_{\beta}$. In our approach the action through Lie derivative is equivalent to the action of $(\text{v}_{\mu}\partial^{\mu}+\d \text{v}_{\mu}q^{\mu})\triangleright \d x_{\beta}=\d \text{v}_{\beta}$ and $[\text{v}_{\mu}\partial^{\mu}+\d \text{v}_{\mu}q^{\mu},\d]=0$. Our approach is more suitable for studying NC case (see section VI.). 

Hidden supersymmetry proposed in \cite{jarvis} could be interpreted as having origin in  additional vector-like Grassmann coordinates. The action of superspace realization of Lorentz generators (\ref{realM}) on physical superfields and possible physical consequences are still under consideration and will be presented elsewhere.

\section{$\kappa$-Minkowski space  in bicrossproduct basis}
In $\kappa$-Minkowski space\footnote{Greek indices $(\mu,\nu,...)$ are  from 0 to 3, and Latin indices $(i,j,...)$ from 1 to 3. Summation over repeated indices is assumed.} with deformed coordinates $\{\hat{x}_{\mu}\}$ we have
\begin{equation}\label{kappa}
[\hat{x}_{i},\hat{x}_{j}]=0, \quad [\hat{x}_{0},\hat{x}_{i}]=ia_{0}\hat{x}_{i},
\end{equation}
where $a_{0}$ is the deformation parameter. The deformed coproducts $\Delta$ for momentum generators $p_{\mu}$ and Lorentz generators $\hat{M}_{\mu\nu}$ in bicrossproduct basis \cite{Majid-Ruegg} are
\begin{equation}\begin{split}\label{bicroscoproduct}
 &\Delta p_{0}=p_{0}\otimes 1+1\otimes p_{0},\quad  \Delta p_{i}=p_{\mu}\otimes 1+e^{a_{0}p_{0}}\otimes p_{i},\\
\Delta \hat{M}_{i0}=&\hat{M}_{i0}\otimes 1+e^{a_{0}p_{0}}\otimes \hat{M}_{i0}-a_{0}p_{j}\otimes \hat{M}_{ij} , \quad  \Delta \hat{M}_{ij}=\hat{M}_{ij}\otimes 1+ 1\otimes \hat{M}_{ij}.\\
\end{split}\end{equation}
The algebra generated by $\hat{M}_{\mu\nu}$ and $p_{\mu}$ is called $\kappa$-Poincar\'{e} algebra where $\hat{M}_{\mu\nu}$ generate undeformed Lorentz algebra, $p_{\mu}$ satisfy $[p_{\mu},p_{\nu}]=0$ and the commutation relations $[\hat{M}_{\mu\nu},p_{\lambda}]$   are given in \cite{Majid-Ruegg}. Equations in (\ref{bicroscoproduct}) describe the coalgebra structure of the $\kappa$-Poincar\'{e} algebra and together with antipode and counit   make the  $\kappa$-Poincar\'{e}-Hopf algebra. We have the action (for more details see \cite{kovacevic-meljanac}) 
$\blacktriangleright\  : \hat{\cal H}(\hat{x},p)\mapsto\hat{\cal A}(\hat{x})$, where $\hat{\cal H}(\hat{x},p)$ is the algebra generated by $\hat{x}_{\mu}$ and $p_{\mu}$ and $\hat{\cal A}(\hat{x})$ is a subalgebra of $\hat{\cal H}(\hat{x},p)$ generated by $\hat{x}_{\mu}$: 
\begin{equation}\begin{split}\label{crnodjelovanje}
&\hat{x}_{\mu} \blacktriangleright \hat{g}(\hat{x})=\hat{x}_{\mu}\hat{g}(\hat{x}),\quad p_{\mu}\blacktriangleright 1=0, \quad \hat{M}_{\mu\nu}\blacktriangleright 1=0 \\
&p_{\mu}\blacktriangleright \hat{x}_{\nu}=-i\eta_{\mu\nu}, \quad  \hat{M}_{\mu\nu}\blacktriangleright \hat{x}_{\lambda}=\eta_{\nu\lambda}\hat{x}_{\mu}-\eta_{\mu\lambda}\hat{x}_{\nu}.\\
\end{split}\end{equation}
Namely, using coproducts (\ref{bicroscoproduct}) and action (\ref{crnodjelovanje}) one can extract the following commutation relations between $\hat{M}_{\mu\nu}$, $p_{\mu}$, and $\hat{x}_{\mu}$:  
\begin{equation}\label{impuls-xkapa}
[p_{0},\hat{x}_{\mu}]=-i\eta_{0\mu}, \quad [p_{k},\hat{x}_{\mu}]=-i\eta_{k\mu}+ia_{\mu}p_{k},
\end{equation}
\begin{equation}\label{MxLie}
[\hat{M}_{\mu\nu},\hat{x}_{\lambda}]=\eta_{\nu\lambda}\hat{x}_{\mu}-\eta_{\mu\lambda}\hat{x}_{\nu}-ia_{\mu}\hat{M}_{\nu\lambda}+ia_{\nu}\hat{M}_{\mu\lambda}.
\end{equation}
 
 The realization corresponding to bicrossproduct basis for $\hat{M}_{\mu\nu}$, $p_{\mu}$, and $\hat{x}_{\mu}$ in terms of undeformed $x_{\mu}$ and $\partial_{\mu}$ is\footnote{Here the superscript $(o)$ denotes that the Lorentz generators and NC coordinates $\hat{x}$ are realized only in terms of undeformed $x_{\mu}$ and $\partial_{\mu}$.} :
 \begin{equation}\begin{split}\label{bicrosrealization}
 \hat{x}^{(o)}_{i}&=x_{i}, \quad \hat{x}^{(o)}_{0}=x_{0}+ia_{0}x_{k}\partial_{k},  \quad  p_{\mu}=-i\partial_{\mu}\\
\hat{M}^{(o)}_{i0}=x_{i}\Bigg(\frac{1-Z}{ia_{0}}+\frac{ia_{0}}{2}\partial^{2}_{k}&-\frac{2}{ia_{0}}\text{sh}^{2}\big(\frac{1}{2}A\big)Z\Bigg)-\big(x_{0}+ia_{0}x_{k}\partial_{k}\big)\partial_{i},\quad \hat{M}^{(o)}_{ij}=x_{i}\partial_{j}-x_{j}\partial_{i},\\
\end{split}\end{equation}
where $A=-ia_{0}\partial_{0}$ and $Z=e^A$  (for more details see \cite{ms06} and \cite{Meljanac-2}).

\section{Bicovariant differential calculus}
In the paper by Sitarz \cite{Sitarz} there is a construction of a bicovariant differential calculus \cite{woro} on $\kappa$-Minkowski space compatible with Lorentz covariance condition (\ref{LcovCon}), but with an extra one-form $\phi$, which transforms as a singlet under the Lorentz generators. The algebra generated by $\hat{x}_{\mu}$ and one-forms $\hat{\xi}_{\mu}$, $\phi$  is closed in one-forms.

The deformed exterior derivative is defined by $\hat{\d}^2 = 0$, $[\hat{\d},\hat{x}_{\mu}]=\hat{\xi}_{\mu}$  and satisfies ordinary Leibniz rule. In the bicovariant calculus it is also assumed that the coproduct for Lorentz generator $\hat{M}_{\mu\nu}$ and momentum generator $p_{\mu}$ is in the bicrossproduct basis (\ref{bicroscoproduct}), the action $\blacktriangleright$\footnote{Sitarz denotes this action with $\triangleright$.} is defined in (\ref{crnodjelovanje}) and it is extended to one-forms by
\begin{equation}\begin{split}\label{djelovanjeSitarz}
&\hat{\xi}_{\mu}\blacktriangleright 1=\hat{\xi}_{\mu}, \quad \phi\blacktriangleright 1=\phi\\
p_{\mu}\blacktriangleright \hat{\xi}_{\nu}=p_{\mu}&\blacktriangleright\phi=\hat{M}_{\mu\nu}\blacktriangleright\phi=0, \quad \hat{M}_{\mu\nu}\blacktriangleright \hat{\xi}_{\lambda}=\eta_{\nu\lambda}\hat{\xi}_{\mu}-\eta_{\mu\lambda}\hat{\xi}_{\nu}. \\ 
\end{split}\end{equation}
From coproducts (\ref{bicroscoproduct}) and eq.(\ref{djelovanjeSitarz}) we can find commutation relations $[\hat{M}_{\mu\nu}, \hat{\xi}_{\lambda}]$ and $[p_{\mu}, \hat{\xi}_{\nu}]$. In addition to Eqs. (\ref{impuls-xkapa}) and (\ref{MxLie}) we have 
\begin{equation}\begin{split}
 [p_{\mu},&\hat{\xi}_{\nu}]=[p_{\mu},\phi]=[\hat{M}_{\mu\nu},\phi]=0,\\
 &[\hat{M}_{\mu\nu},\hat{\xi}_{\lambda}]=\eta_{\nu\lambda}\hat{\xi}_{\mu}-\eta_{\mu\lambda}\hat{\xi}_{\nu}.\\
 \end{split}\end{equation}
The Lorentz covariance condition 
\begin{equation}\begin{split}\label{LcovCon}
\hat{M}_{\mu\nu}\blacktriangleright & \hat{f}(\hat{x},\hat{\xi})\hat{g}(\hat{x},\hat{\xi})=m\big(\Delta \hat{M}_{\mu\nu}\blacktriangleright\hat{f}\otimes \hat{g}\big)\\
&\hat{M}_{\mu\nu}\blacktriangleright \hat{\text{d}}\hat{f}=\hat{\text{d}}(\hat{M}_{\mu\nu}\blacktriangleright \hat{f})\\
\end{split}\end{equation}
  implies 
\begin{equation}\label{Md=0}
[\hat{M}_{\mu\nu},\hat{\text{d}}]=0,
\end{equation}
where $\hat{\d} \hat{f}=\hat{\d}\blacktriangleright \hat{f}=[\hat{\d}, \hat{f}]\blacktriangleright 1$ and $\hat{\d}\blacktriangleright 1=0$.

Sitarz claims that the algebra\footnote{The correspondence between algebra in \cite{Sitarz} and (\ref{closed})  is $\frac{1}{\kappa}=-ia_{0}$, $x_{\mu}=\hat{x}_{\mu}$, $\d x_{\mu}=\hat{\xi}_{\mu}$, $\phi=\phi$, $N_{i}=\hat{M}_{0i}$, and $M_{i}=\epsilon_{ijk}\hat{M}_{jk}$.} between one-forms $\hat{\xi}_{\mu}$, $\phi$ and NC coordinate $\hat{x}_{\mu}$ that is compatible with (\ref{LcovCon}) - (\ref{Md=0}) is given by
\begin{equation}\begin{split}\label{closed}
&[\hat{x}_{\mu},\phi]=\hat{\xi}_{\mu}, \quad [\hat{x}_{0},\hat{\xi}_{0}]=-a^{2}_{0}\phi,\\
[\hat{x}_{i},\hat{\xi}_{j}]=-ia_{0}&\delta_{ij}\big(\hat{\xi}_{0}+ia_{0}\phi\big), \quad [\hat{x}_{0},\hat{\xi}_{i}]=0, \quad [\hat{x}_{i},\hat{\xi}_{0}]=-ia_{0}\hat{\xi}_{i}.\\
\end{split}\end{equation}

 The realization for $\hat{x}_{\mu}$ is given in (\ref{bicrosrealization}) and the realization for one-forms $\hat{\xi}_{\mu}$ and $\phi$ that satisfies (\ref{closed}) can be given in terms of undeformed $x_{\mu}$, $\partial_{\mu}$ and $\xi_{\mu}$\footnote{The algebra of  undeformed operators is defined in Section II. and for $\blacktriangleright$ action we have $x_{\mu}\blacktriangleright 1=\hat{x}_{\mu}$, $\xi_{\mu}\blacktriangleright 1=\hat{\xi}_{\mu}$, $\partial_{\mu}\blacktriangleright 1=0$ and $q_{\mu}\blacktriangleright 1=0$. }. The realizations\footnote{For more details see \cite{MKJj}} for one-forms and exterior derivative are
\begin{equation}\begin{split}\label{realizatSitarz}
&\hat{\xi}_{0}=\xi_{0}\Big(1+\frac{a^{2}_{0}}{2}\square\Big)+ia_{0}\xi_{k}\partial_{k}, \quad \hat{\xi}_{k}=\xi_{k}-ia_{0}\xi_{0}\partial_{k}Z^{-1},\\
\phi=-&\hat{\text{d}}_{s}=\xi_{0}\Big(\frac{Z^{-1}-1}{ia_{0}}+\frac{ia_{0}}{2}\square\Big)-\xi_{k}\partial_{k},\quad \square=\partial^{2}_{i}Z^{-1}-\frac{4}{a^{2}_{0}}\text{sh}^{2}\big(\frac{1}{2}A\big),\\
\end{split}\end{equation}
where we have denoted the exterior derivative for Sitarz's case with $\hat{\text{d}}_{s}$.

Relations (\ref{closed}), (\ref{realizatSitarz}) and (\ref{djelovanje}) imply $\phi=-\hat{\text{d}}_{s}$  and $\phi\triangleright 1=0$, so that the algebra\footnote{The algebra $\hat{{\cal SA}}$ is generated by $\hat{x}_{\mu}$, $\hat{\xi}_{\mu}$ and $\phi$, and the algebra ${\cal SA}^{\star }$ is generated by $x_{\mu}$ and $\xi_{\mu}$ but with $\star$-multiplication. The star-product is defined by $f(x,\xi)\star g(x,\xi)=\hat{f}(\hat{x},\hat{\xi})\hat{g}(\hat{x},\hat{\xi})\triangleright 1$. } $\hat{{\cal SA}}$ is not isomorphic to ${\cal SA}^{\star }$.
Also the problem with this construction is that the Lorentz generators $\hat{M}_{\mu\nu}$ cannot be realized in terms of $x_{\mu}$, $\partial_{\mu}$, $\xi_{\mu}$ and $q_{\mu}$ in order to satisfy  Lorentz covariance condition (\ref{LcovCon}) which implies (\ref{Md=0}). Namely, if we take the realization (\ref{realizatSitarz}) for $\hat{\text{d}}_{s}$ and just want to find realization for $\hat{M}_{\mu\nu}$ so that $[\hat{M}_{\mu\nu},\hat{\text{d}}_{s}]=0$ is fulfilled, then these $\hat{M}_{\mu\nu}$ do not  satisfy the Lorentz algebra (\ref{lorentz}). 
 
 Furthermore, in \cite{MKJj}  differential algebras $\mathfrak{D}_{1}$ and $\mathfrak{D}_{2}$ of classical dimension were constructed (avoiding the extra form $\phi$), where  all conditions were satisfied, except (\ref{MxLie}) and $\hat{M}_{\mu\nu}$ does not commute with exterior derivative. All these arguments lead to a conclusion that for the fixed realization (\ref{bicrosrealization}) for $\hat{x}_{\mu}$, there is no realization for $\hat{M}_{\mu\nu}$ that satisfies $\kappa$-Poincar\'{e}-Hopf algebra (\ref{bicroscoproduct}) and Lorentz covariance condition (\ref{LcovCon}), (\ref{Md=0}).

\section{$\kappa$-Minkowski spacetime from twist and NC one-forms}
In this section we will construct the noncommutative coordinates $\hat{x}_{\mu}$, coproducts, and NC one-forms using the twist operator.

\subsection{$\kappa$-Minkowski spacetime from twist}
 We start with an Abelian twist (see \cite{Govindarajan-1}, \cite{klry08}, \cite{klry09} and \cite{Borowiec-3})
\begin{equation}
{\cal F} =\exp\bigl( -A\otimes x_{k}\partial_{k}\bigr),
\label{1}
\end{equation}
where $A=ia\partial=-ia_{0}\partial_{0}$. The bidifferential operator \eqref{1} satisfies all the properties of a twist
 (2-cocycle condition and normalization) and leads to noncommutative coordinates
\begin{equation}
\hat{x}_{\mu}=m_{0}\Bigl( {\cal F}^{-1}\triangleright(x_{\mu}\otimes id) \Bigr). \label{1.5}
\end{equation}
 It follows that for this twist we get a realization for $\hat{x}_{\mu}$ exactly as in (\ref{bicrosrealization}).
The twist given by Eq. \eqref{1} also leads to an associative star product
\begin{equation}\label{starprodukt}
f(x)\star g(x)=m_{0} \Bigl( {\cal F}^{-1}\triangleright(f\otimes g) \Bigr). 
\end{equation}

If we define the operators $M_{\mu\nu}$ as $M_{\mu\nu}=x_{\mu}\partial_{\nu}-x_{\nu}\partial_{\mu}$, then $M_{\mu\nu}$ generate the undeformed Lorentz algebra, but their coproducts, obtained from the twist \eqref{1} do not close in the Poincar\'{e}-Hopf algebra. For this reason we consider the algebra $\mathfrak{igl}(4)$, generated by  $\partial_{\mu}$ and $L_{\mu\nu}=x_{\mu}\partial_{\nu}$, which also has a Hopf algebra structure \cite{Kovacevic222}. The coproducts of $\partial_{\mu}$ and $L_{\mu\nu}$ calculated as $\Delta \partial_{\mu}={\cal F}\Delta_{0}\partial_{\mu}{\cal F}^{-1}$, and analogously for $L_{\mu\nu}$, are
\begin{eqnarray}
&& \Delta\partial_{0}=\Delta_{0}\partial_{0}, \quad \Delta \partial_{i} =\partial_{i}\otimes 1+e^{A}\otimes \partial_{i} \label{8} \\
&& \Delta L_{ij}=\Delta_{0}L_{ij}, \quad \Delta L_{00}=\Delta_{0}L_{00}+A\otimes L_{kk} \label{9} \\
&& \Delta L_{i0}=L_{i0}\otimes 1+e^{-A}\otimes L_{i0}, \quad \Delta L_{0i}=L_{0i}\otimes 1+e^{A}\otimes L_{0i}-ia_{0}\partial_{i}\otimes L_{kk}. \label{12}
\end{eqnarray}
Coproducts of the momenta $p_{\mu}$, obtained from \eqref{8} by expressing $p_{\mu}$ in terms of $\partial_{\mu}$ ($p_{\mu}= -i\partial_{\mu}$), coincide with the coproducts of momenta in the bicrossproduct basis (\ref{bicroscoproduct}) (see also \cite{Majid-Ruegg}). On the other hand, coproducts of the Lorentz generators $M_{\mu\nu}$, calculated from Eqs. \eqref{9} and \eqref{12} as $\Delta M_{\mu\nu} =\Delta L_{\mu\nu} -\Delta L_{\nu\mu}$, are different from the ones in the bicrossproduct basis (\ref{bicroscoproduct}) 
(more precisely, $\Delta M_{i0}$ is different), \cite{Kovacevic222}. In this way we have constructed the  $\mathfrak{igl}(4)$-Hopf algebra structure using twist $\mathcal{F}$.

We point out that in \cite{rmatrix, mali} it is shown that for  Lorentz generators in bicrossproduct basis (\ref{bicrosrealization}) the twist $\mathcal{F}$ gives the correct Hopf algebra structure (\ref{bicroscoproduct}) of $\kappa$-Poincar\'{e} algebra generated by $\hat{M}_{\mu\nu}$ and $p_{\mu}$.

\subsection{Noncommutative  one-forms from twist}

Our aim is to construct an exterior derivative $\hat{\d}$ and noncommutative one-forms $\hat{\xi}_{\mu}$ with the following properties:
\begin{eqnarray}
&& \hat{\d}^{2}=0, \quad [\hat{\d},\hat{x}_{\mu}]=\hat{\xi}_{\mu} \label{15} \\
&& \{ \hat{\xi}_{\mu},\hat{\xi}_{\nu} \}=0, \quad [\hat{\xi}_{\mu},\hat{x}_{\nu}]=K_{\mu \nu}^{\lambda}\hat{\xi}_{\lambda}, \ \ K_{\mu \nu}^{\lambda} \in {\mathbb C} \label{16} \\
&& [\hat{\xi}_{\mu}, \hat{x}_{\nu}] -[\hat{\xi}_{\nu}, \hat{x}_{\mu}]=ia_{\mu}\hat{\xi}_{\nu}-ia_{\nu}\hat{\xi}_{\mu} \quad \text{(consistency \ condition)}, \label{17}
\end{eqnarray}
where we have introduced $a_{\mu}=(a_{0},\vec{0})$ so that (\ref{kappa}) can be written in a unified way as
\begin{equation}
[\hat{x}_{\mu},\hat{x}_{\nu}]=i(a_{\mu}\hat{x}_{\nu}-a_{\nu}\hat{x}_{\mu}).
\end{equation}

We want to find a realization of NC one-forms in terms of the undeformed algebra ${\cal SH}(x,\partial,\xi,q)$.
If we now calculate $\hat{\xi}_{\mu}$, by analogy to Eq. \eqref{1.5}, as $\hat{\xi}_{\mu}=m_{0} \Bigl( {\cal F}^{-1}\triangleright(\xi_{\mu}\otimes id) \Bigr)$, we get $\hat{\xi}_{\mu}=\xi_{\mu}$, so that the LHS of \eqref{17} equals  $0$, while the RHS gives $ia_{\mu}\hat{\xi}_{\nu}-ia_{\nu}\hat{\xi}_{\mu}$ and the consistency condition is not fulfilled. Obviously we need to extend the twist defined in (\ref{1}). 

We have shown that the bicovariant differential calculus  \'{a} la Sitarz \cite{Sitarz} could not be realized in terms of  Heisenberg or super-Heisenberg algebra. In the next section we will propose a new version of bicovariant calculus compatible with $\mathfrak{igl}(4)$-Hopf algebra.

\section{Extended twist}
Our main goal is to construct a twist so that our bicovariant calculus satisfies the following properties:
\begin{enumerate}
\item The bicovariant calculus has classical dimension, i.e. there is no extra form like $\phi$.
\item The algebra between $\hat{\xi}_{\mu}$ and $\hat{x}_{\mu}$ is closed in one-forms.
\item Generators $M_{\mu\nu}$ satisfy the Lorentz algebra.
\item The condition $[M_{\mu\nu}, \hat{\d}]=0$, which is sufficient condition for (\ref{LcovCon}), (\ref{Md=0}).
\end{enumerate}
In order to satisfy all the requirements for $\hat{\xi}_{\mu}$ and $\hat{\d}$ we define the extended twist
\begin{equation}
{\cal F}_{ext}=\exp\Bigl( -A\otimes (x_{k}\partial_{k}+\xi_{k}q_{k}) \Bigr).
\end{equation}
This twist leads to
\begin{eqnarray}\label{xext}
&& \hat{x}_{i}=m_{0}\Bigl( {\cal F}_{ext}^{-1}\triangleright(x_{i}\otimes id) \Bigr)=x_{i}, \quad \hat{x}_{0}=m_{0}\Bigl( {\cal F}_{ext}^{-1}\triangleright(x_{0}\otimes id) \Bigr) =x_{0}+ia_{0}(x_{k}\partial_{k}+\xi_{k}q_{k}) \\
&& \hat{\xi}_{\mu}=m_{0}\Bigl( {\cal F}_{ext}^{-1}\triangleright(\xi_{\mu}\otimes id) \Bigr) =\xi_{\mu}. \label{24}
\end{eqnarray}
Although the realization of $\hat{x}_{0}$ is changed with the addition of a term containing Grassmann variables, $\hat{x}_{\mu}$ still satisfy the same commutation relations Eq. (\ref{kappa}), but the commutation relations between $\hat{x}_{\mu}$ and $\hat{\xi}_{\mu}$ are no longer all equal to $0$
\begin{equation}
[\hat{\xi}_{\mu},\hat{x}_{i}]=0, \quad [\hat{\xi}_{0},\hat{x}_{0}]=0, \quad [\hat{\xi}_{i},\hat{x}_{0}]=-ia_{0}\hat{\xi}_{i}. \label{25}
\end{equation}
Inserting \eqref{25} into \eqref{17} shows that in this case the consistency condition and the requirement from \eqref{16} are satisfied. Note that $\hat{x}_{\mu}$, $\hat{\xi}_{\mu}$, $\partial_{\mu}$ and $q_{\mu}$ generate the deformed super-Heisenberg algebra ${\cal \hat{SH}}$,which also has super-Hopf-algebroid structure.

We now want to introduce an exterior derivative $\hat{\d}$ such that \eqref{15} is also fulfilled and gives rise to the same expression for $\hat{\xi}_{\mu}$ as \eqref{24}. It is easily shown that this is achieved with
\begin{equation}\label{exd}
\hat{\d}=\xi^{\alpha}\partial_{\alpha}=\d.
\end{equation}

For the exterior derivative $\d$ in (\ref{exd}), we wanted to constructed an operator $\hat{M}_{\mu\nu}=\hat{M}^{(0)}_{\mu\nu}+\text{Grassmann part}$ by extending the realization in $\mathcal{SH}$ with the property that it commutes with exterior derivative, i.e. $[\hat{M}_{\mu\nu}, \d]=0$, but in doing so, we find that this operator does not satisfy the Lorentz algebra (\ref{lorentz}). Hence, exterior derivative (\ref{exd}) is not be compatible with $\kappa$-Poincar\'{e}-Hopf algebra in the bicrossproduct basis even if we consider the realizations in $\mathcal{SH}$.

Since $\hat{\xi}_{\mu}$ are undeformed, their $\triangleright$ and $\blacktriangleright$ actions are the same as for $\xi_{\mu}$. Our construction can be extended to forms of higher order in a natural way. E.g., the space of two-forms can be defined as the space generated by $\hat{\xi}_\mu \wedge \hat{\xi}_\nu$. These two-forms then automatically satisfy
\begin{equation}
\hat{\xi}_\mu \wedge \hat{\xi}_\nu= \xi_\mu \wedge \xi_\nu = -\xi_\nu \wedge \xi_\mu= -\hat{\xi}_\nu \wedge \hat{\xi}_\mu.
\end{equation}

Now we define the extended $\star$-product with
\begin{equation}\label{26}
f(x,\xi)\star g(x,\xi) = m_{0} \Bigl( {\cal F}_{ext}^{-1}\triangleright f\otimes g \Bigr). 
\end{equation} 
For $f(x)$ and $g(x)$, functions of $x$ only, \eqref{26} coincides with (\ref{starprodukt}), and if $f(\xi)$ and $g(\xi)$ are functions of $\xi$ only, their extended $\star$-product is just the ordinary multiplication, $f(\xi)\star g(\xi)=f(\xi)g(\xi)$. As before (see (\ref{starprodukt})), the extended $\star$-product can be equivalently defined with the $\triangleright$ action: $f(x,\xi)\star g(x,\xi) =\hat{f}(\hat{x},\hat{\xi}) \hat{g}(\hat{x},\hat{\xi}) \triangleright 1$.

In order to get a Lorentz i.e. $\mathfrak{igl}(4,\mathbb{R})$ covariant action,  generators of $\mathfrak{gl}(4)$ also need to be extended. $L_{\mu\nu}^{ext}$ are defined by $L_{\mu\nu}^{ext}=x_{\mu}\partial_{\nu}+\xi_{\mu}q_{\nu}$. It can be easily checked that $L_{\mu\nu}^{ext}$, defined in this way, still satisfy the $\mathfrak{igl}(4)$ algebra, and furthermore, that they commute with $\d$, $[L_{\mu\nu}^{ext},\d]=0$, so that
\begin{equation}
L_{\mu\nu}^{ext}\blacktriangleright \hat{\xi}_{\lambda}=[L_{\mu\nu}^{ext},\hat{\xi}_{\lambda}]\blacktriangleright 1=\d[L_{\mu\nu}^{ext},\hat{x}_{\lambda}]\blacktriangleright 1=\d (L_{\mu\nu}^{ext}\blacktriangleright \hat{x}_{\lambda})=\hat{\xi}_{\mu}\eta_{\nu\lambda}. \label{27}
\end{equation}
The results for the coproducts, obtained from the extended twist, are
\begin{eqnarray}
&& \Delta q_{0}=\Delta_{0}q_{0}, \quad \Delta q_{i}=q_{i}\otimes 1+(-)^{deg}e^{A}\otimes q_{i} \\
&& \Delta L_{ij}^{ext}=\Delta_{0}L_{ij}^{ext}, \quad \Delta L_{00}^{ext} =\Delta_{0}L_{00}^{ext} +A\otimes L_{kk}^{ext} \label{31} \\
&& \Delta L_{i0}^{ext} =L_{i0}^{ext}\otimes 1+e^{-A}\otimes L_{i0}^{ext}, \quad \Delta L_{0i}^{ext} =L_{0i}^{ext}\otimes 1+e^{A}\otimes L_{0i}^{ext}-ia_{0}\partial_{i}\otimes L_{kk}^{ext}. \label{34}
\end{eqnarray}
The coproducts of $\partial_{0}$ and $\partial_{i}$, calculated in the same way, are again given by Eq. \eqref{8}.

The action of $L_{\mu\nu}^{ext}$ on the product of $\hat{x}_{\rho}$ and $\hat{\xi}_{\sigma}$, calculated in three different ways:
\\ (i) $L_{\mu\nu}^{ext}\blacktriangleright \hat{x}_{\rho}\hat{\xi}_{\sigma} =[L_{\mu\nu}^{ext},\hat{x}_{\rho}]\blacktriangleright \hat{\xi}_{\sigma} +\hat{x}_{\rho}\blacktriangleright ([L_{\mu\nu}^{ext},\hat{\xi}_{\sigma}]\blacktriangleright 1)$
\\ (ii) $L_{\mu\nu}^{ext}\blacktriangleright \hat{x}_{\rho}\hat{\xi}_{\sigma} =\bigl(L_{\mu\nu(1)}^{ext} \blacktriangleright \hat{x}_{\rho}\bigr)\bigl(L_{\mu\nu(2)}^{ext} \blacktriangleright \hat{\xi}_{\sigma}\bigr)$
\\ (iii) $L_{\mu\nu}^{ext}\blacktriangleright \hat{x}_{\rho}\hat{\xi}_{\sigma} =\bigl(L_{\mu\nu(1)}^{ext} \blacktriangleright \hat{x}_{\rho}\bigr)\d \bigl(L_{\mu\nu(2)}^{ext} \blacktriangleright \hat{x_{\sigma}}\bigr)$,
\\ gives the same result, i.e., the action is in accordance with bicovariant calculus. One can easily show that neither the last equality, (iii), nor \eqref{27} would be satisfied had we used the ordinary definition of $L_{\mu\nu}$ ($L_{\mu\nu}=x_{\mu}\partial_{\nu}$). Similar expressions can be written in terms of the $\triangleright$ action and the extended $\star$ product.
Hence, we have constructed  $\mathfrak{igl}(4)$-Hopf algebra structure using the extended twist $\mathcal{F}_{ext}$ 
that satisfies all the requirements for bicovariant calculus (listed in the beginning of this section 1-4). Note that the coproduts of $M_{\mu\nu}=L^{ext}_{\mu\nu}-L^{ext}_{\nu\mu}$ calculated in this way differ from the one in bicrossproduct basis.

The action \eqref{27} could also be obtained using the ordinary definition of $L_{\mu\nu}$ and promoting these generators to Lie derivatives. Using Cartan's identity, one would get
\begin{equation}
\mathcal{L}_{x_\mu \partial_\nu} \triangleright \xi_\lambda = \d(\mathcal{L}_{x_\mu \partial_\nu} \triangleright x_\lambda) = \d(x_\mu \eta_{\nu\lambda} )= \xi_\nu \eta_{\nu\lambda}.
\end{equation}
However, there is a problem in this approach. Namely, promoting $L_{\mu\nu}$ to Lie derivatives would again give the realization (\ref{bicrosrealization}) for $\hat{x}_{\mu}$ and in the case of deformed one-forms it would give $\hat{\xi}_\mu = \xi_\mu$, i.e., the consistency condition would not be fulfilled.

\section{Outlook and discussion}

We have shown that if the NC coordinates (\ref{kappa}) are given only in terms of Heisenberg algebra ${\cal H}(x,\partial)$, then
there is no realization of Lorentz generators  compatible with all the requirements of bicovariant calculus \cite{Sitarz}.  Hence, if one wants to unify $\kappa$-Minkowski spacetime, $\kappa$-Poincar\'{e} algebra and differential forms it is crucial to embed them into $\kappa$-deformed super-Heisenberg algebra $\hat{{\cal SH}}(\hat{x},\hat{\xi},\partial,q)$. This is explicitly done for $\kappa$-deformed $\mathfrak{igl}(4)$-Hopf algebra using the extended twist. We choose a bicrossproduct basis just as one example, but similar constructions for other bases are also possible.

In section VI we have constructed a bicovariant differential calculus compatible with $\kappa$-deformed $\mathfrak{igl}(4)$-Hopf algebra. The question is whether it is possible to construct bicovariant calculus compatible with $\kappa$-Poincar\'{e}-Hopf algebra. One has to develop the notion of superphase space and its super-Hopf-algebroid structure. The idea is to generalize the method developed in \cite{mali} where we have analyzed  the quantum phase space, its deformation and Hopf-algebroid structure. In \cite{mali} we have also constructed  $\kappa$-Poincar\'{e}-Hopf algebra from twist.

  We present the realization of $\kappa$-Poincar\'{e} algebra compatible with bicovariant differential calculus in the bicrossproduct basis. Starting with the extended realization of $\hat{x}_{\mu}$, 
\begin{equation}
 \hat{x}_{i}=x_{i}, \quad \hat{x}_{0}=x_{0}+ia_{0}(x_{k}\partial_{k}+\xi_{k}q_{k}) 
\end{equation}
 and demanding the Lorentz algebra (\ref{lorentz}), (\ref{bicroscoproduct}), and (\ref{MxLie}) we find the realization for the Lorentz generators:
 \begin{equation}\begin{split}
 &\hat{M}_{ij}= x_{i}\partial_{j}-x_{j}\partial_{i}+\xi_{i}q_{j}-\xi_{j}q_{i}\\
 \hat{M}_{i0}=\hat{M}^{(0)}_{i0}&+\xi_{i}(q_{0}Z^2 +ia_{0}q_{k}\partial_{k})-[\xi_{0}q_{i}+ia_{0}(\xi_{k}q_{k}\partial_{i}+\xi_{k}\partial_{k}q_{i})]
 \end{split}\end{equation}
 where $\hat{M}^{(0)}_{i0}$ is given in (\ref{bicrosrealization}). The requirement that the Lorentz generators $\hat{M}_{\mu\nu}$ commutes with  exterior derivatives $\tilde{\d}$,  i.e. $[\tilde{\d},\hat{M}_{\mu\nu}]=0$ is fulfilled for
 \begin{equation}
 \tilde{\d}=\frac{1}{1+\frac{a^{2} _{0}}{2}\Box}\left[\xi_{0}\left(\frac{\text{sh}A}{ia_{0}}+\frac{ia_{0}}{2}\partial^{2}_{i}Z^{-1}\right)+\xi_{j}\partial_{j}Z      \right]
 \end{equation}
The corresponding one-forms are defined as $\tilde{\xi}_{\mu}=[\tilde{\d}, \hat{x}_{\mu}]$. We are working on generalizing the results in \cite{mali} in order to construct the extended twist operator within the superphase space which will provide the correct $\kappa$-Poincar\'{e}-Hopf algebra compatible with bicovariant calculus.
All the details of this construction will be presented elsewhere.

Our main motivation for studying these problems is related to the fact that the general theory of relativity together with uncertainty principle leads to NC spacetime. In this setting, the notion of smooth spacetime geometry  and its symmetries  are generalized using the  Hopf algebraic approach. Further development  of the approach presented in this paper  will lead to possible application to the construction of NC quantum field theories (especially electrodynamics and gauge theories), quantum gravity models, particle statistics, and modified dispersion relation.\\

\bigskip

\noindent{\bf Acknowledgment}\\
We would like to thank Peter Schupp for useful comments.  This work was supported by the Ministry of Science and Technology of the Republic
of Croatia under contract No. 098-0000000-2865. R.\v{S}. gratefully acknowledges support from the DFG within the Research Training Group 1620 ``Models of Gravity''.\\

\end{document}